\documentstyle[psfig,subfigure]{mn}

\newif\ifAMStwofonts

\ifoldfss
  \ifCUPmtlplainloaded \else
    \NewTextAlphabet{textbfit} {cmbxti10} {}
    \NewTextAlphabet{textbfss} {cmssbx10} {}
    \NewMathAlphabet{mathbfit} {cmbxti10} {} 
    \NewMathAlphabet{mathbfss} {cmssbx10} {} 
  \fi
  \ifAMStwofonts
    \ifCUPmtlplainloaded \else
      \NewSymbolFont{upmath} {eurm10}
      \NewSymbolFont{AMSa} {msam10}
      \NewMathSymbol{\upi}     {0}{upmath}{19}
      \NewMathSymbol{\umu}     {0}{upmath}{16}
      \NewMathSymbol{\upartial}{0}{upmath}{40}
      \NewMathSymbol{\leqslant}{3}{AMSa}{36}
      \NewMathSymbol{\geqslant}{3}{AMSa}{3E}

    \fi
  \fi
\fi 

\ifnfssone
  \newmathalphabet{\mathit}
  \addtoversion{normal}{\mathit}{cmr}{m}{it}
  \addtoversion{bold}{\mathit}{cmr}{bx}{it}
  \newmathalphabet{\mathbfit} 
  \addtoversion{normal}{\mathbfit}{cmr}{bx}{it}
  \addtoversion{bold}{\mathbfit}{cmr}{bx}{it}
  \newmathalphabet{\mathbfss} 
  \addtoversion{normal}{\mathbfss}{cmss}{bx}{n}
  \addtoversion{bold}{\mathbfss}{cmss}{bx}{n}
  \ifAMStwofonts
    \ifCUPmtlplainloaded \else
      \UseAMStwoboldmath
      \makeatletter
      \new@mathgroup\upmath@group
      \define@mathgroup\mv@normal\upmath@group{eur}{m}{n}
      \define@mathgroup\mv@bold\upmath@group{eur}{b}{n}
      \edef\UPM{\hexnumber\upmath@group}
      \new@mathgroup\amsa@group
      \define@mathgroup\mv@normal\amsa@group{msa}{m}{n}
      \define@mathgroup\mv@bold\amsa@group{msa}{m}{n}
      \edef\AMSa{\hexnumber\amsa@group}
      \makeatother
      \mathchardef\upi="0\UPM19
      \mathchardef\umu="0\UPM16
      \mathchardef\upartial="0\UPM40
      \mathchardef\leqslant="3\AMSa36
      \mathchardef\geqslant="3\AMSa3E
    \fi
  \fi
\fi 

\ifnfsstwo
  \DeclareMathAlphabet{\mathbfit}{OT1}{cmr}{bx}{it}
  \SetMathAlphabet\mathbfit{bold}{OT1}{cmr}{bx}{it}
  \DeclareMathAlphabet{\mathbfss}{OT1}{cmss}{bx}{n}
  \SetMathAlphabet\mathbfss{bold}{OT1}{cmss}{bx}{n}
  \ifAMStwofonts
    \ifCUPmtlplainloaded \else
      \DeclareSymbolFont{UPM}{U}{eur}{m}{n}
      \SetSymbolFont{UPM}{bold}{U}{eur}{b}{n}
      \DeclareSymbolFont{AMSa}{U}{msa}{m}{n}
      \DeclareMathSymbol{\upi}{0}{UPM}{"19}
      \DeclareMathSymbol{\umu}{0}{UPM}{"16}
      \DeclareMathSymbol{\upartial}{0}{UPM}{"40}
      \DeclareMathSymbol{\leqslant}{3}{AMSa}{"36}
      \DeclareMathSymbol{\geqslant}{3}{AMSa}{"3E}
    \fi
  \fi
\fi 

\ifCUPmtlplainloaded \else
  \ifAMStwofonts \else 
    \def\upi{\pi}
    \def\umu{\mu}
    \def\upartial{\partial}
  \fi
\fi

\title[]{Searching for tidal tails - investigating galaxy harassment}
\author[J. I. Davies et al.]{J. I. Davies$^{1}$, S. Roberts$^{1}$ and S. Sabatini$^{2}$ \\
$^{1}$School of Physics and Astronomy, Cardiff University, The Parade, Cardiff, CF24 3YB, UK. \\ 
$^{2}$Observatorio Astronomico di Roma, via Frascati 33, I-00040, Monte Porzio, Italy.
}

\begin{document}

\date{Accepted 000000. Received 000000; in original form 000000}

\pagerange{\pageref{firstpage}--\pageref{lastpage}} \pubyear{2003}

\maketitle

\label{firstpage}

\begin{abstract}
Galaxy harassment has been proposed as a physical process that morphologically transforms low surface density disc galaxies into dwarf elliptical galaxies in clusters. It has been used to link the observed very different morphology of distant cluster galaxies (relatively more blue galaxies with 'disturbed' morphologies) with the relatively large numbers of dwarf elliptical galaxies found in nearby clusters. One prediction of the harassment model is that the remnant galaxies should lie on low surface brightness tidal streams or arcs. We demonstrate in this paper that we have an analysis method that is sensitive to the detection of arcs down to a surface brightness of 29 $B\mu$ and then use this method to search for arcs around 46 Virgo cluster dwarf elliptical galaxies. We find no evidence for tidal streams or arcs and consequently no evidence for galaxy harassment as a viable explanation for the relatively large numbers of dwarf galaxies found in the Virgo cluster.
\end{abstract}


\section{Introduction}
 Cold Dark Matter (CDM) hierarchical galaxy and large scale structure formation models predict many more small dark matter halos than can be associated with known dwarf galaxies (Moore et al., 1999a, Klypin et al., 1999). Methods of avoiding this discrepency within the models usually involve the global suppression of star formation in small halos, either by the prevention of infalling gas and/or the early expulsion of gas (Dekel \& Silk, 1986, Efstathiou, 1992). These two plausible physical mechanisms are critically challenged because the  lack of large numbers of dwarf galaxies is environmentally dependent. In the Local Group the ratio of dwarf ($-14<M_{B}<-10$) to giant ($-19>M_{B}$) galaxies (the Dwarf to Giant Ratio or DGR) is about 5 (Mateo, 1998). A similar value has been found by us for galaxies in other relatively low density environments (Roberts et al., 2004a). The DGR is very different in the much higher density environment of the Virgo cluster, being about 20 (Sabatini et al., 2003, Roberts et al. 2004b)). There is good evidence to suggest that other dense environments also have relatively large numbers of dwarf galaxies (Kambas et al., 2000, Bernstein et al., 1995). 

In a hierarchical model a cluster such as Virgo is constructed from infalling galaxies, like the Milky Way and its companions, but the constituents of the cluster now are very different to that expectation. So, within the current paradigm, the cluster environment must have significantly affected its galaxy population in a way that does not affect galaxies in the field. This point has, of course, been known for many years because the centres of clusters are where we predominately find elliptical galaxies which, according to hierarchical models, are assembled from the merger of gas rich disc galaxies. Distant clusters also have a higher fraction of very blue galaxies compared to the field, again showing that the cluster environment is having an effect on the way galaxies evolve (Butcher \& Oemler, 1984, Dressler et al., 1994).  What is still not clear though is just why there are so many more dwarf galaxies per giant in these rich environments. This has led to a number of ideas about how the cluster environment may either make more dark matter halos become visible in the form of dwarf galaxies or how it may lead to the creation of dwarf galaxies. 

Galaxy 'squelching' is one proposed mechanism for making cluster dark matter halos more visible (Tully et al., 2002). The basic idea is that dwarf galaxies in a large overdensity (that will eventually form a cluster like Virgo) form early, before re-ionisation, so that they can retain gas and form stars. Those outside of proto-clusters form later and the gas is then too hot to fall into small dark matter halos. Tully et al. say that there is 'qualitative' agreement between their model and observation, but this was before the WMAP result pushed the re-ionisation epoch back to $z=20$ (Spergel et al., 2003), making the idea much less convincing. Babul \& Rees (1992) have proposed that in the cluster environment the expulsion of gas by dwarf galaxies is inhibited by the intra-cluster gas. This pressure confinement means that cluster dwarf galaxies are much more likely to retain gas to form subsequent populations of stars. Although a viable physical mechanism, we have shown (Sabatini et al., 2004) that this mechanism is only effective within the very central regions of a cluster like Virgo, where the intra-cluster gas density is high enough. But, this is also the region where dwarf galaxies would be totally destroyed by tidal forces (Sabatini et al., 2004). Given also the competition between pressure confinement and ram pressure stripping it is not clear whether this is actually a viable mechanism at all even if dense gas were to exist outside of the cluster core.

An alternative scenario is galaxy 'harassment'. This is a process that creates dwarf galaxies in the cluster environment by the transformation of low surface density disc galaxies. Numerical simulations by Moore et al., (1996, 1999b) indicate that the numerous high speed gravitational interactions that affect low density discs as they move through the cluster environment can remove the outer disc stars and 'fatten up' the disc. The result is very much like the numerous dwarf elliptical (dE) galaxies that we see in clusters. Moore et al., have proposed that the 'excess' dE galaxy population of clusters like Virgo is due to this morphological transformation of low surface density disc galaxies. It is clear that tidal interactions of some form or another must play an important role in cluster galaxy evolution. Observations of distant clusters clearly show many apparently tidally disrupted galaxies (Dressler et al., 1994). The origin of intra-cluster stars (Ferguson et al., 1998) and planetary nebulae (Feldmeier et al., 1998) are presumably the result of such interactions. 
In summary there is good evidence for tidal interactions (Moore et al. 1998), but is the evidence for morphological transformation so strong ? 

In this paper we look at one of the predictions of the harassment model to scrutinise it in the light of new observational data and data reduction techniques. The harassment (morphogical transformation) model predicts that stars removed from the harassed galaxies will orbit in 'prominent' narrow streams that follow the orbital path of the galaxy. The galaxy itself is a brightening on this stellar stream. A rough calculation predicts that an annulus of stars tidally stripped from a $M_{B}=-20$ (Moore et al., 1996) galaxy distributed around a radius of 1 Mpc (of order the size of a galaxy cluster) and thickness 2 kpc (of order the size of a dwarf galaxy) would have a surface brightness of about 27.5 $B\mu$ ($B\mu$ - blue magnitudes per sq arc sec). Features like this, that almost certainly have a tidal origin, have been found in a number of clusters. For example Calcaneo-Roldan et al., (2000) describe CCD observations of a low surface brightness arc in the Centaurus cluster (originally discovered by David Malin). The arc has a length of about 100 kpc and a width of 2 kpc, its surface brightness is 27.8 $B\mu$. Trentham \& Mobasher (1998) describe an arc in the Coma cluster that is about 80 kpc long and has a surface brightness of order 26.5 $B\mu$. Gregg and West (1998) have identified other possible tidal features in the Coma cluster core that have surface brightnesses of order 27.5 $B\mu$. Though convincingly tidal in origin none of them are convincingly assiciated with a dE galaxy. But, the observations do indicate that if we want to search for arcs as signatures of harassment we need to be sensitive to features that have surface brightnesses as low as 28 $B\mu$. 

Should we find such features in a cluster like Virgo ? According to Calcaneo-Roldan et al., (2000) these tidal features should persist for a few cluster crossing times, which for Virgo is a few billion years (Trentham et al., 2001). Moore et al., (1996) say that the dramatic change in cluster galaxy morphology due to galaxy harassment has occured from $z\approx0.4$ to today. This is a period of 4.6 billion years (standard flat $\Omega_{m}=0.3$, $\Delta_{o}=0.7$ and $H_{o}=70$ km s$^{-1}$ Mpc$^{-1}$ cosmology). Thus our test, as described below, should be sensitive to galaxies that have been harassed over the latter half of the morphological transformation epoch.

\begin{figure*}
\Large
{\bf a)}
\normalsize
\subfigure{\psfig{figure=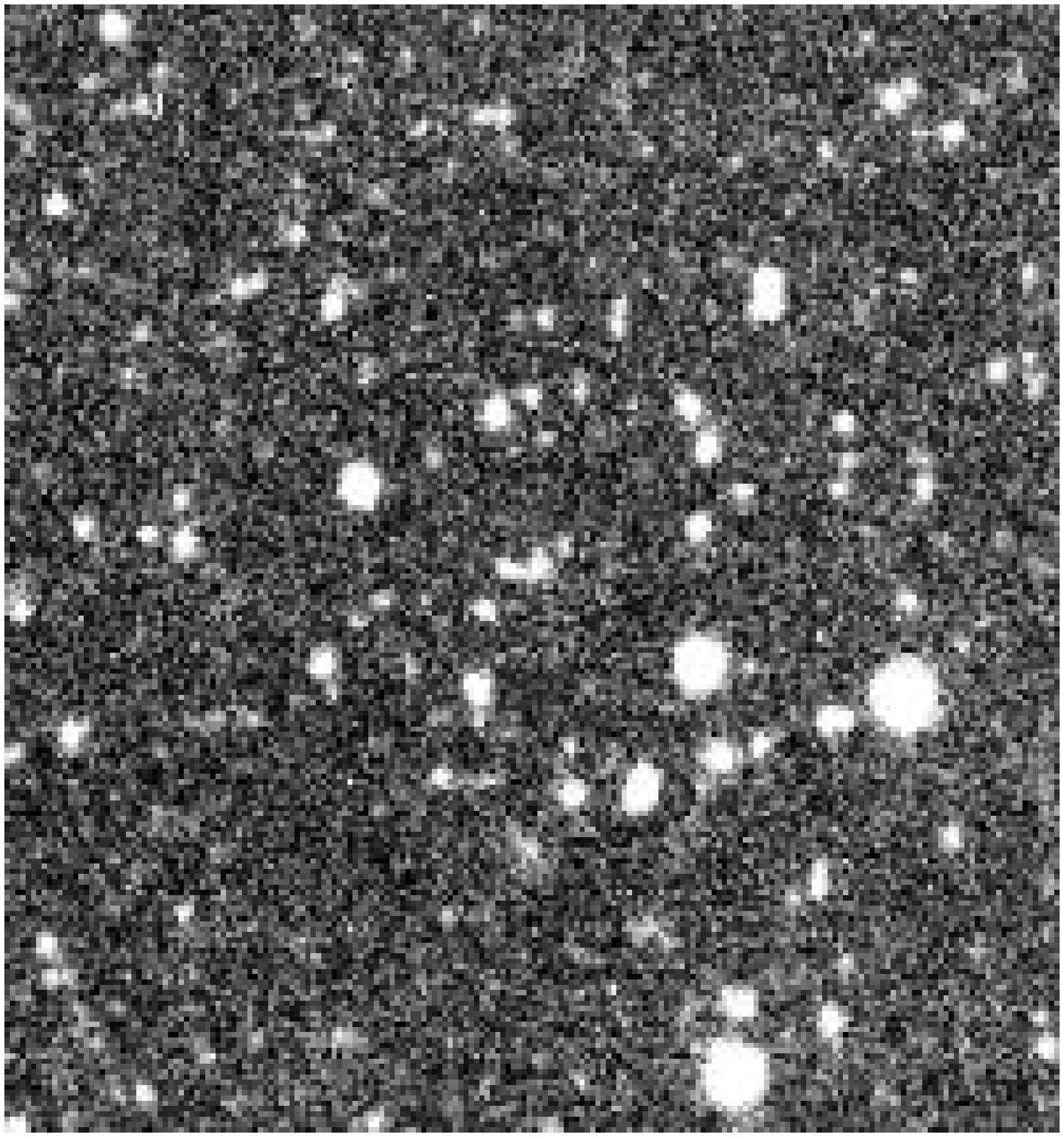,width=6cm}}
\subfigure{\psfig{figure=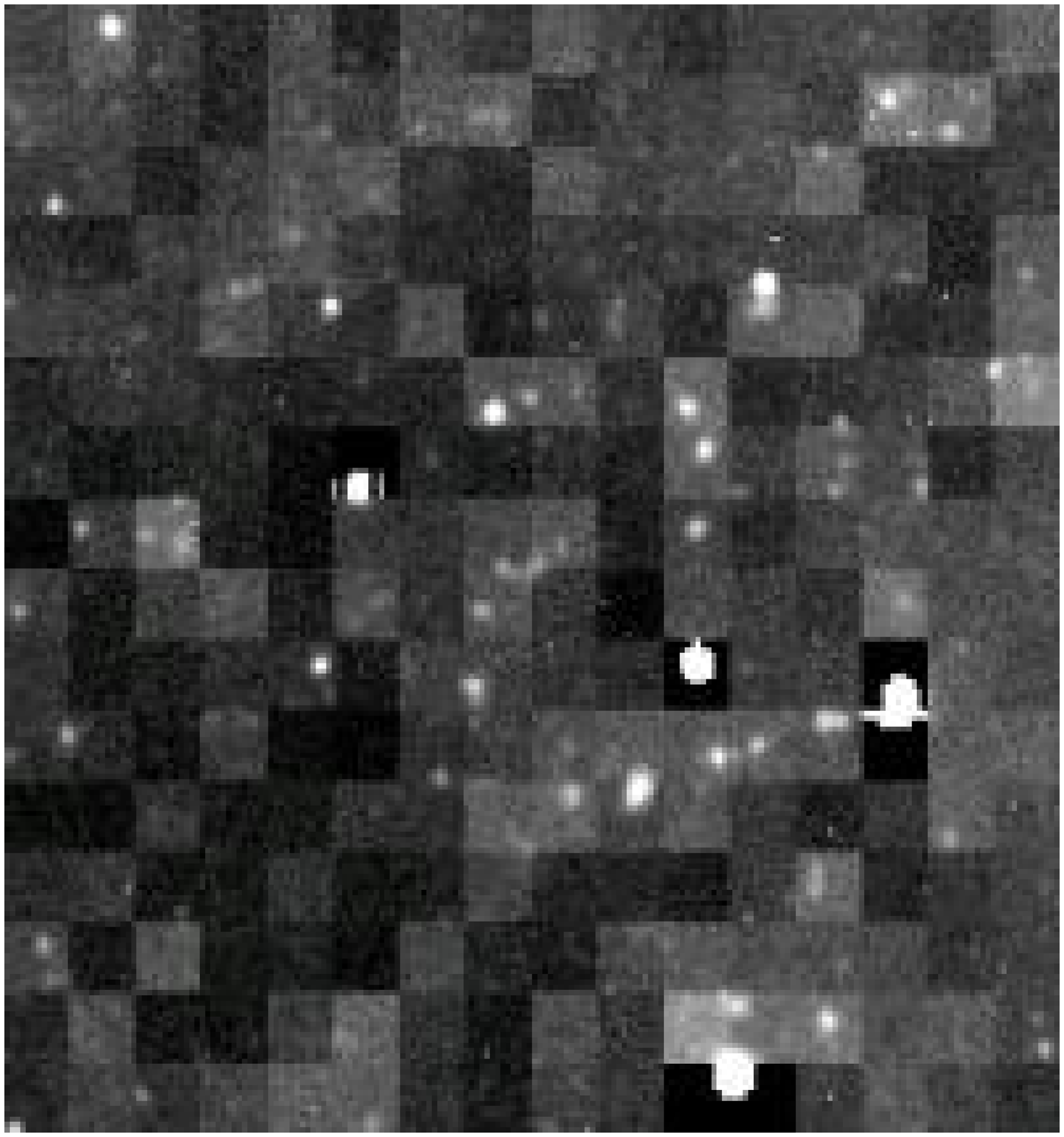,width=6cm}}

\Large
{\bf b)}
\normalsize
\subfigure{\psfig{figure=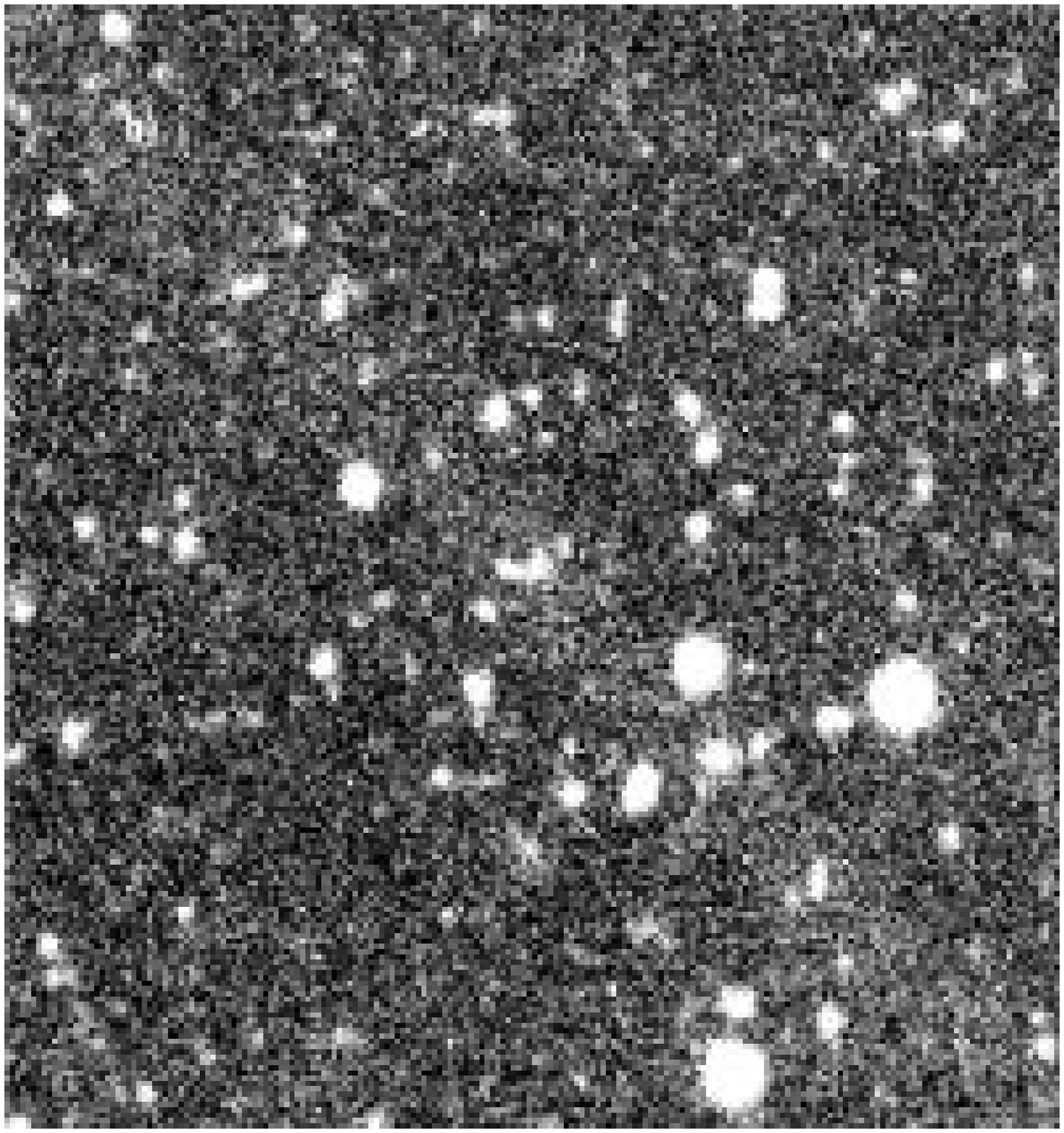,width=6cm}}
\subfigure{\psfig{figure=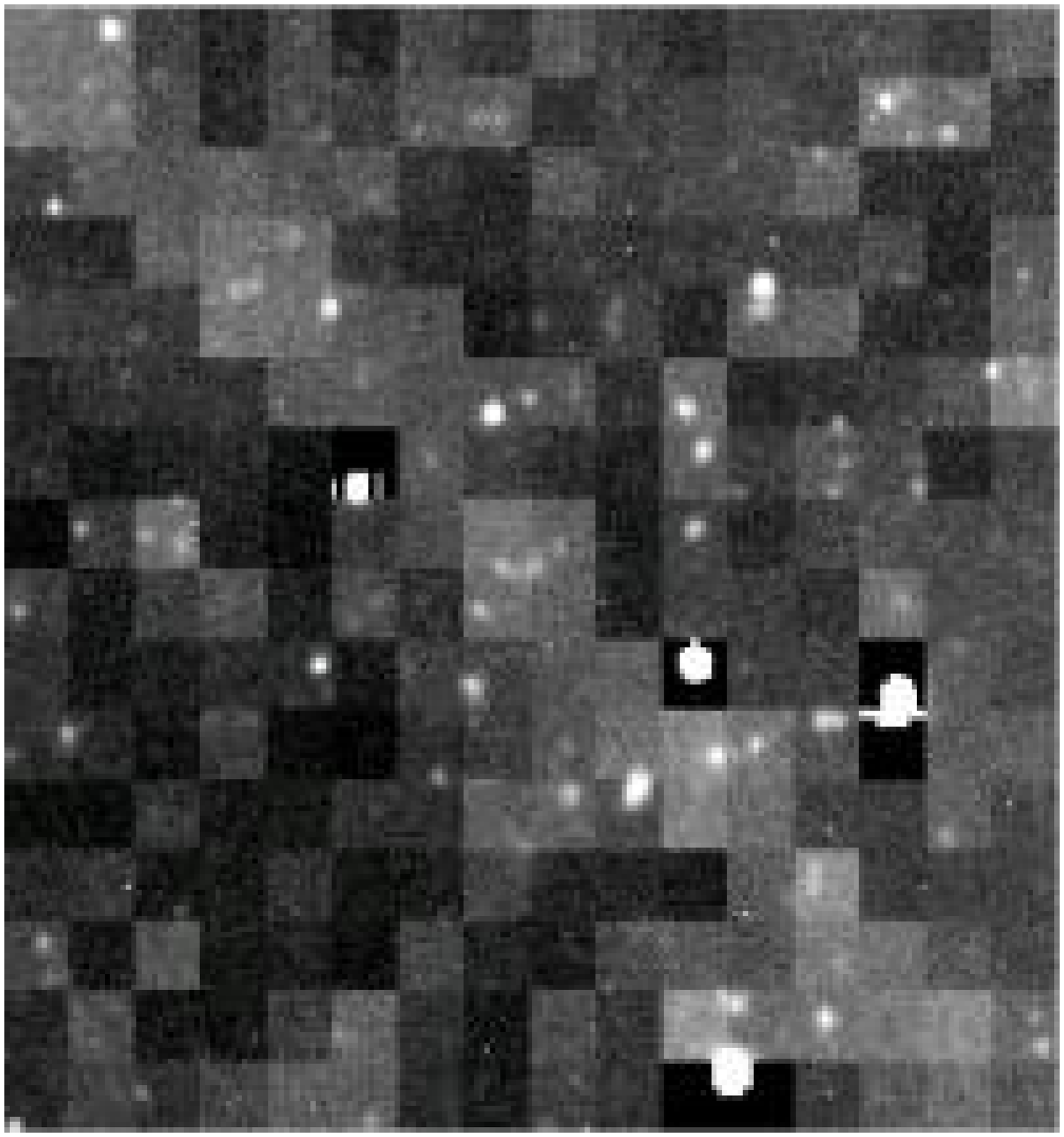,width=6cm}}

\Large
{\bf c)}
\normalsize
\subfigure{\psfig{figure=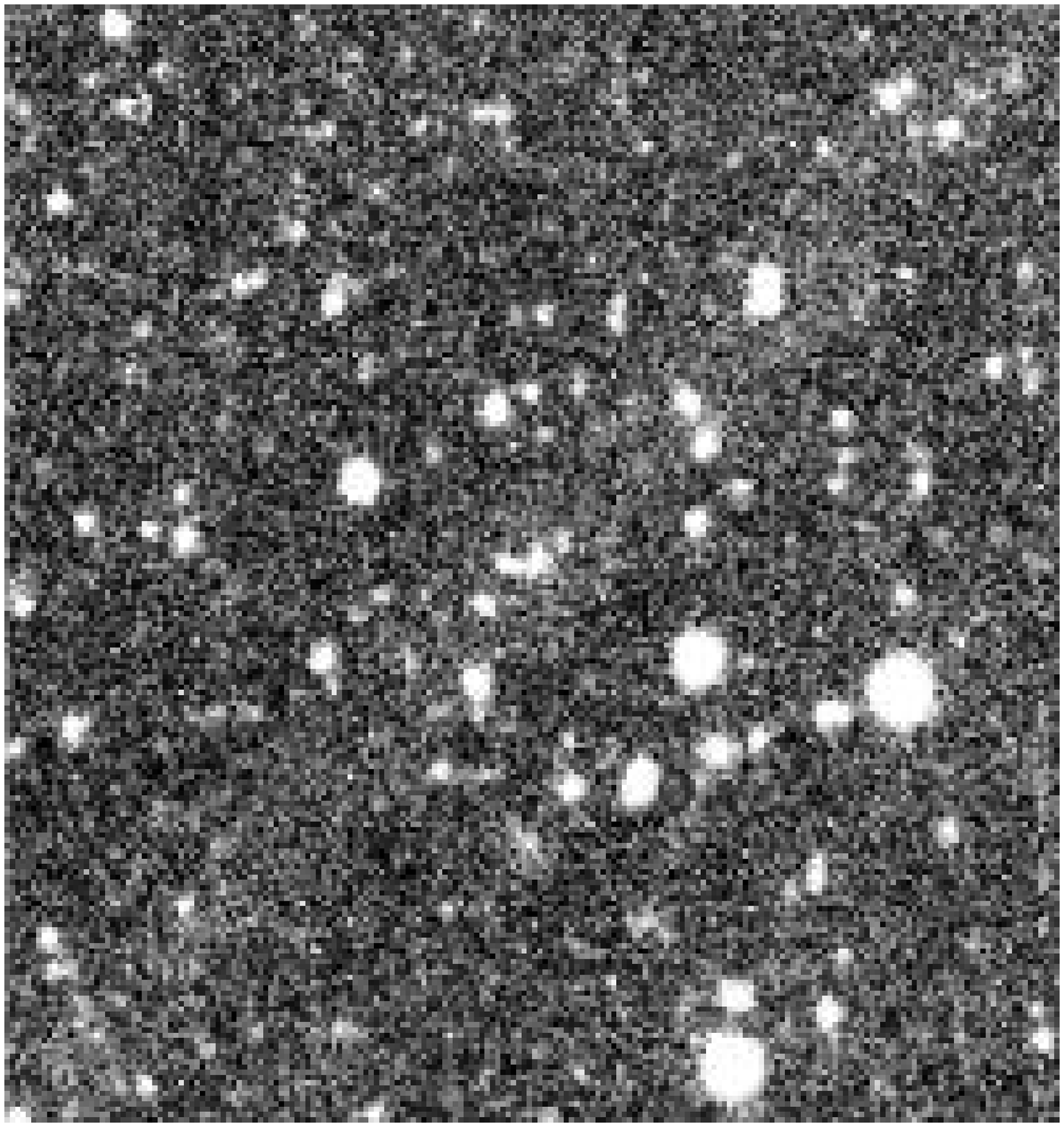,width=6cm}}
\subfigure{\psfig{figure=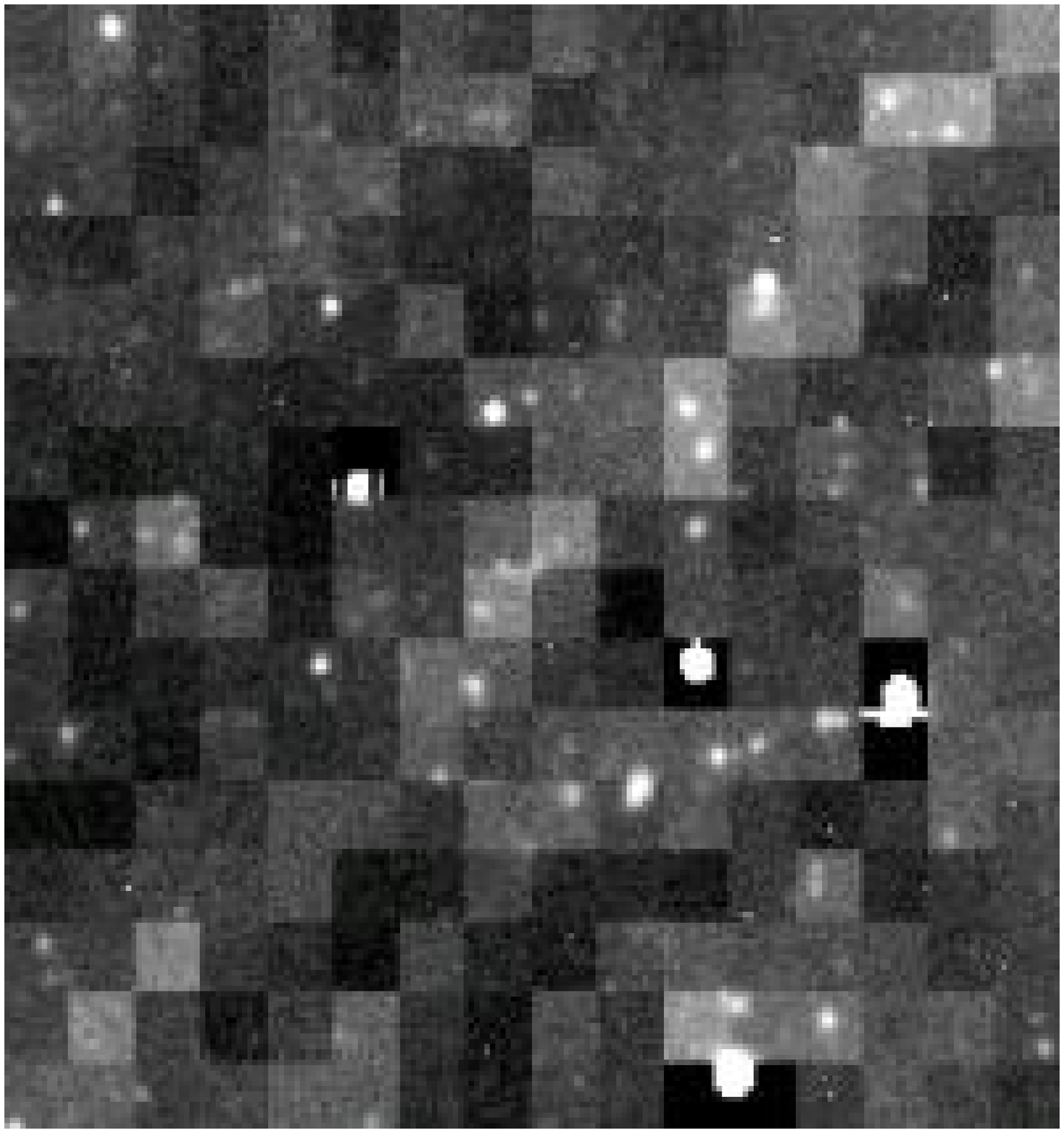,width=6cm}}
\caption
{Data with simulated arc (left) and after Haar smoothing  (right). a) No arcs added, b) Diagonal line (top left to bottom right), c) Diagonal line (bottom left to top right), d)Horizontal line, e) Vertical line. Each is displayed between 1$\sigma$ below sky level to 3$\sigma$ above sky level ($\sigma$ is the sky standard deviation). }
\end{figure*}

\begin{figure*}
\Large
{\bf d)}
\normalsize
\subfigure{\psfig{figure=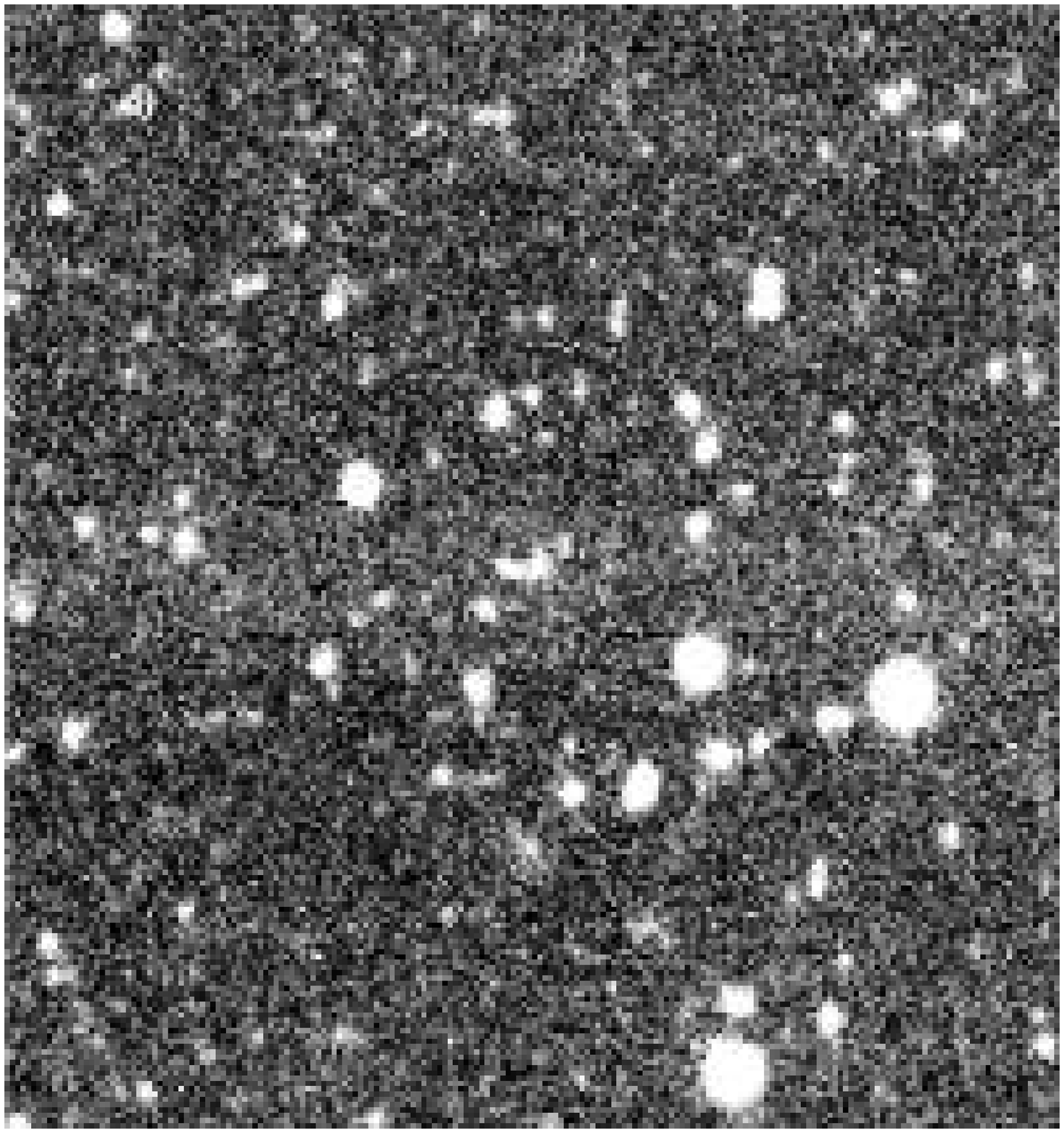,width=6cm}}
\subfigure{\psfig{figure=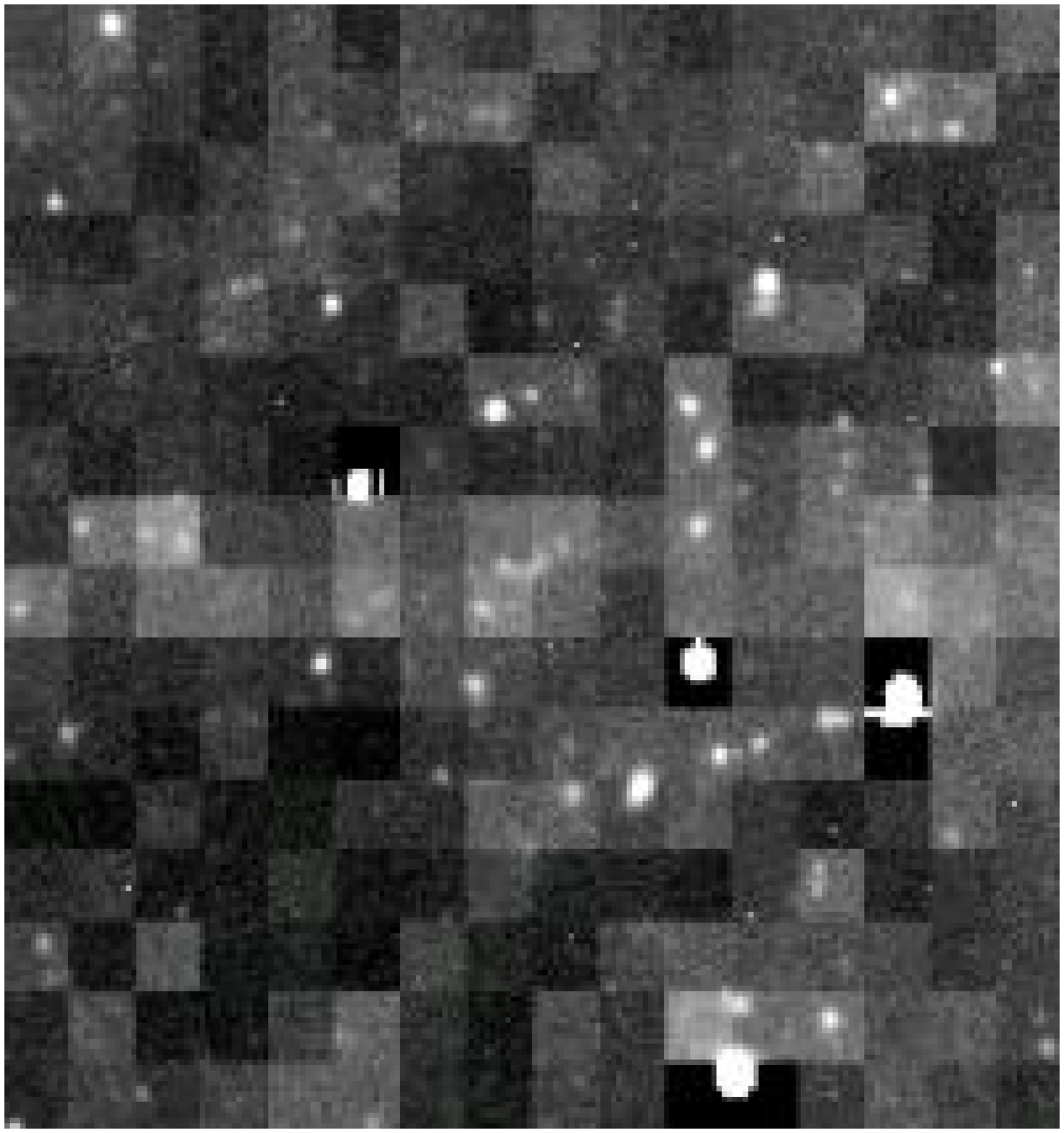,width=6cm}}

\Large
{\bf e)}
\normalsize
\subfigure{\psfig{figure=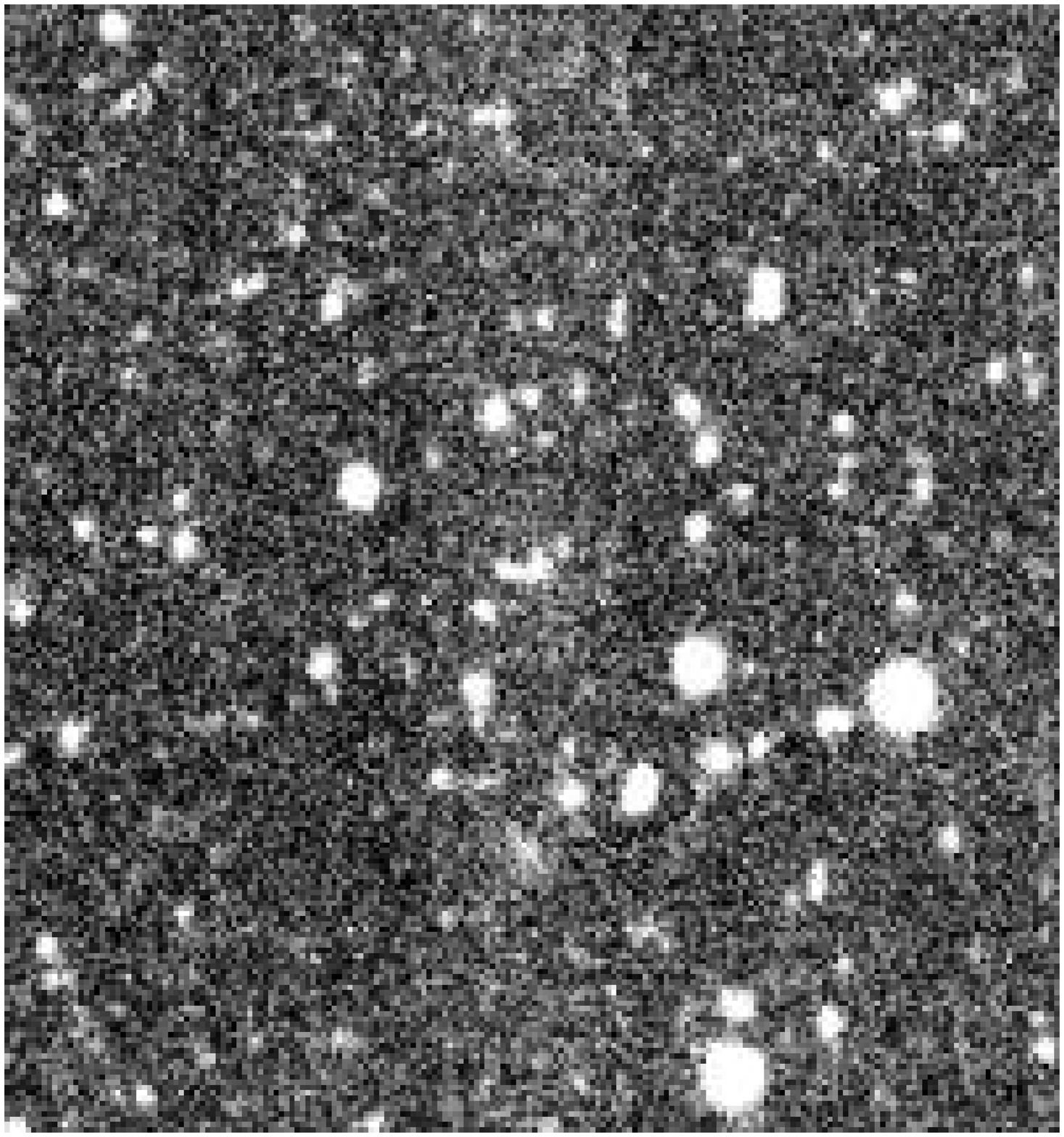,width=6cm}}
\subfigure{\psfig{figure=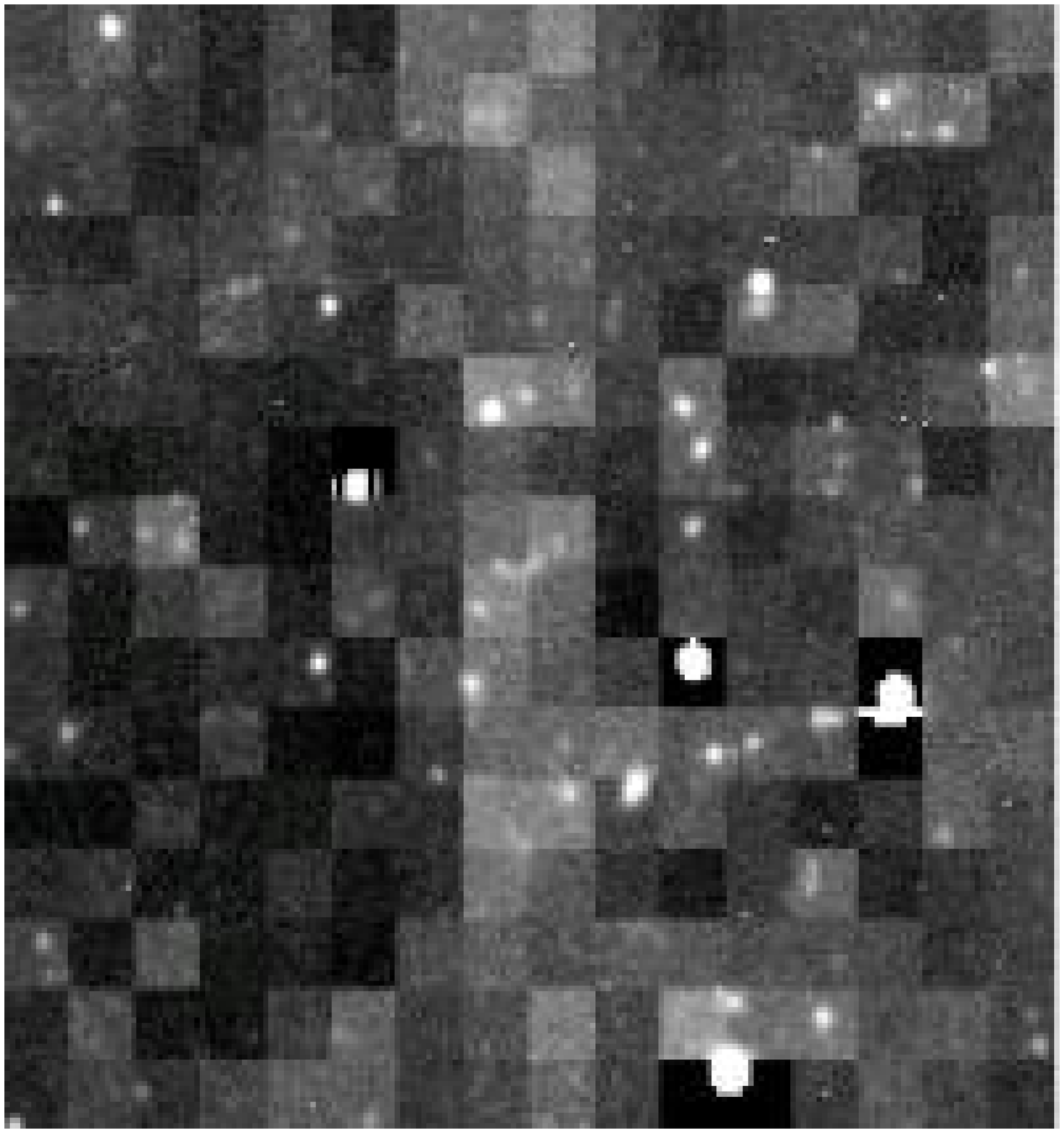,width=6cm}}

 \end{figure*}

\section{Data and data processing}
We have used B band CCD data from the Issac Newton Telescope (INT) Wide Field Camera survey of the Virgo cluster to search for evidence of tidal streams. These data are fully described in Sabatini et al., (2003) and Roberts et al., (2004a). As the basis of our search for tidal features we have used 38 galaxies listed as dE in Roberts et al., (2004b). Of these galaxies 24 (63\%) are also listed in the Virgo Cluster Catalogue (VCC) (Binggeli et al. 1985) as cluster dE galaxies. The typical one sigma sky noise for these frames is $26$ $B\mu$ some 2 magnitudes brighter than required to detect tidal arcs.

To increase the surface brightness limit to 28 $B\mu$ we have smoothed the data. Smoothing can be problematic when looking for Low Surface Brightness  (LSB) features as excessive smoothing, for example with a Gaussian or tophat function, makes all objects appear to be LSB. Ideally, one would like to smooth the 'sky' (those regions that contain no resolved foreground or background objects) without smearing out light from the resolved objects. One way of doing this is to apply the Haar transform. In its simplest form it can be used to smooth over a $2\times2$ pixel area. Instead of storing in memory the 4 pixel values  4 different numbers are stored, from which the original image can be reconstructed - the mean value of the four pixels, a measure of the x-gradient, a measure of the y gradient and a measure of the curvature (for an example see Aboufadel and Sclicker, 1999). If the measures of the gradients and curvatures are as expected from a separate measurement of the background pixel-to-pixel fluctuations (they are just sums of differences of pixel values) then they can be set to zero or reduced. Where the gradients are high (when the pixels are part of a resolved object) the differences can be retained. After transforming back to pixel values, the sky has been smoothed without spreading out the light from resolved objects. To generalise, we form the orthonormal matrix $T$ which by definition has unit magnitude and satisfies $T^{T}=T^{-1}$. If the $2^{n}\times2^{n}$ matrix $X$ contains the pixel values then we can form the matrix 

\begin{equation}
Y=TXT^{T}
\end{equation}

in the $2\times2$ case the matrix $Y$ contains the mean value and gradients described above. The matrix $Y$ can now be filtered as described above and then transformed back using

\begin{equation}
X=T^{T}YT
\end{equation}

For the purpose of this experiment we want to filter on kpc scales (width of the tidal features) and so we bin the data into 1 arc sec pixels and then filter with a matrix $T$ of dimension $16\times16$ (16 arc sec is $\approx$1 kpc at the distance of Virgo, 16 Mpc). In the case of a $16\times16$ matrix, $Y$ contains one element that is the sum of the pixel values and the others are the sums of the differences between pixels - so again we can see if this is consistent with being sky noise and filter as appropriate. An example of the effect this has on a data frame is shown in Fig. 1a, it is clear that the brighter compact objects have not been smoothed. The filtering does tend to break the image up into $16\times16$ blocks, but we will show below that it still enables us to detect features at surface brightness levels of 28 $B\mu$.

To test the method we have added 'arcs' of surface brightness (with random noise) of 28 $B\mu$ to data frames from the INT. We have extracted regions of $256\times256$ 1 arc sec pixels centred on our sample of dE galaxies. This corresponds to about 20 kpc at the distance of the Virgo cluster (same linear size as the arc described in Conselice and Gallagher, 1999). We have chosen this distance because we want the 'arcs' to be almost linear over the region of the data we use. An 'arc' in the plane of the sky of radius 1 Mpc would only deviate from a straight line by one pixel over this distance. The arc described in Calcaneo-Roldan et al., 2000, deviates by 3-4 arc sec over 100 arc sec - so we are expecting almost linear features. Our method (see below) would only then not work if we were unlucky enough to be viewing the galaxy close to the tangent point (orbital velocity projected along the line of sight) of an almost edge-on arc (orbit inclined at $\approx 90^{o}$ to the plane of the sky). The arc would then more resemble a 'tail' as it would extend to one side only of the galaxy. This will occur when the galaxy is less than about one data frame size away from the tangential point. For a 1 Mpc radius orbit this will amount to about 10\% of the time. So, less than 10\% of our target arcs might go undetected because of the detection method we are using (see below). This is an upper limit because not all orbits will be sufficiently inclined. We also expect that the 'arcs' will both be brighter and stable over longer periods closer to the galaxy. In Fig. 1(b-e). we show on the left a random piece of sky with faint 28 $B\mu$ 2 kpc wide arcs added. On the right we show the result of Haar smoothing.

Looking at Fig. 1. it is not really clear that the Haar smoothing has helped very much in making the tidal features more visible. Below we will show that it has, we just need to quantify the effect. Given that our frames are centred on a dE galaxy and that we expect the arcs to be straight lines in our images we can now quantify our detections. After numerous trials, we consider only those pixels (in the filtered image) that have intensity values that lie between the sky background value and the sky background plus 3$\sigma$ ($\sigma$ is the background standard deviation in the filtered frame). We then calculate the angle that a vector connecting the pixel to the centre of the frame (position of the galaxy) makes with the horizontal. In Fig. 2. we show the  distribution of angles derived from the simulated data shown in Fig. 1. Fig. 2a) is for the smoothed data frame with no 'arc' added (a random area of sky) - no significant peak (see below) is seen. Fig 2b) is for a frame that has an 'arc' added, but has not been smoothed - again no significant peak is seen. Figs 2c-f. each shows a significant peak at the angle of the 'arcs' shown in fig. 1. We have quantified the significance of the detection by calculating the standard deviation of the distribution of angles (excluding the peak value) and comparing this to the number in the peak minus the mean value (the Signal-to-Noise ($SN$)). For the histograms shown in Fig 2a and b we find $SN<2$, for those in Fig 2c-f $SN>3$. To summarise, we make no detections in a frame on a random piece of sky (checked with 20 other random pieces of sky) or on frames that have had 'arcs' added, but have not been smoothed. All simulated arcs are detected with a $SN>3$ when smoothed. We conclude that on real data frames a value of $SN<2$ corresponds to a non-detection while $SN>3$ is a definite detection. For $2<SN<3$ the question remains open.

\begin{figure*}
\Large
{\bf a)}
\normalsize
\subfigure{\psfig{figure=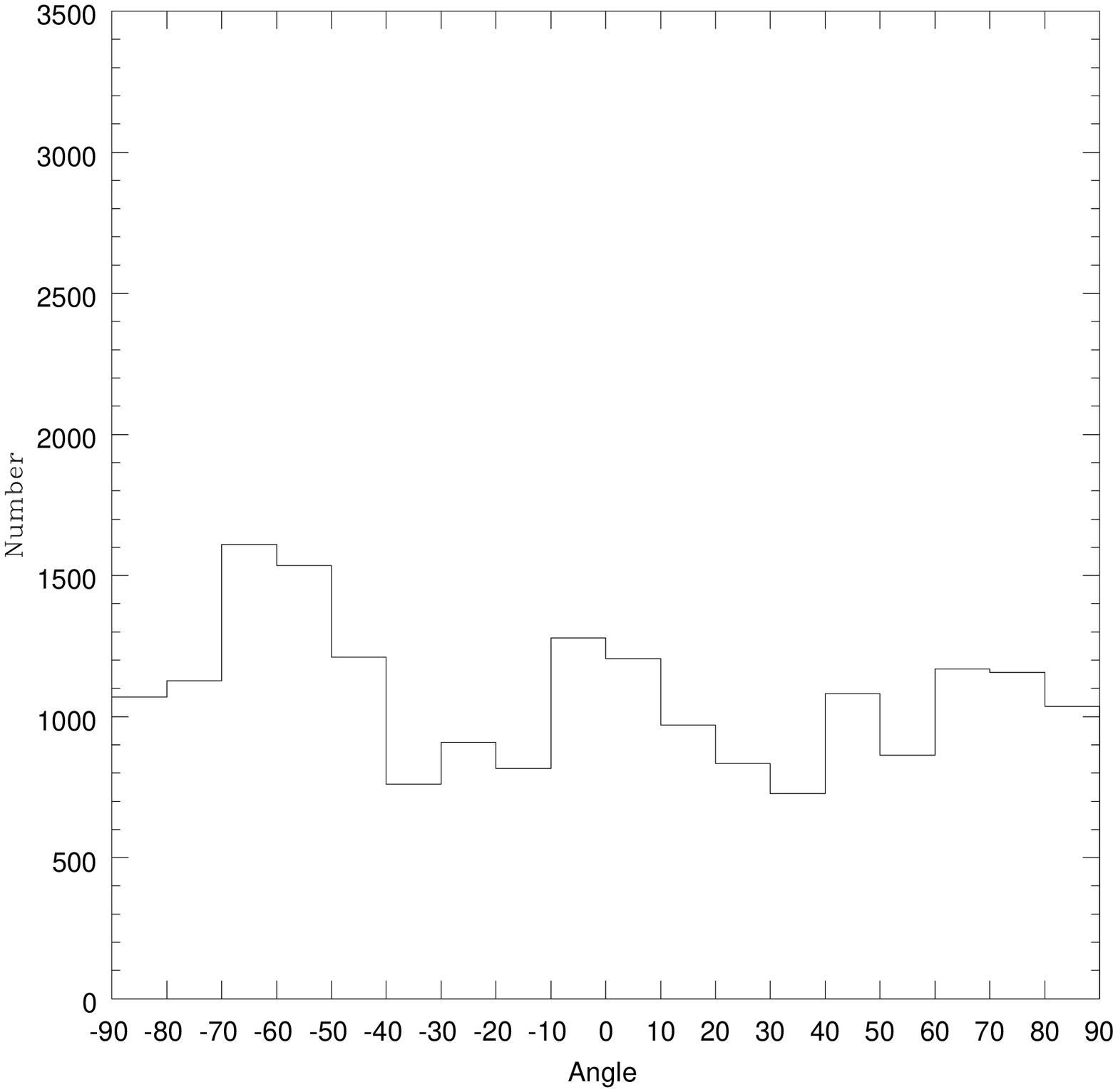,width=6cm}}
\Large
{\bf b)}
\normalsize
\subfigure{\psfig{figure=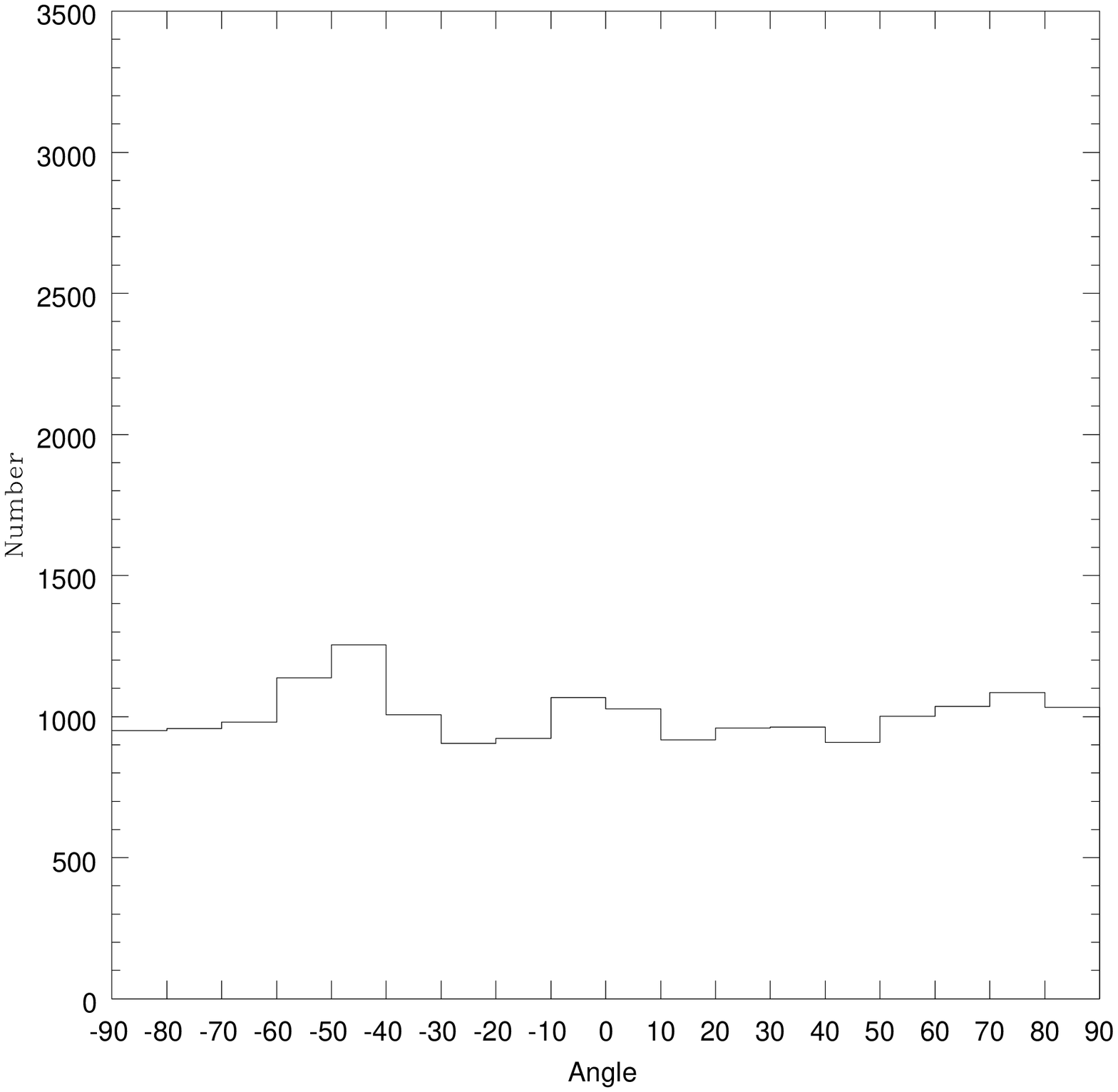,width=6cm}}

\Large
{\bf c)}
\normalsize
\subfigure{\psfig{figure=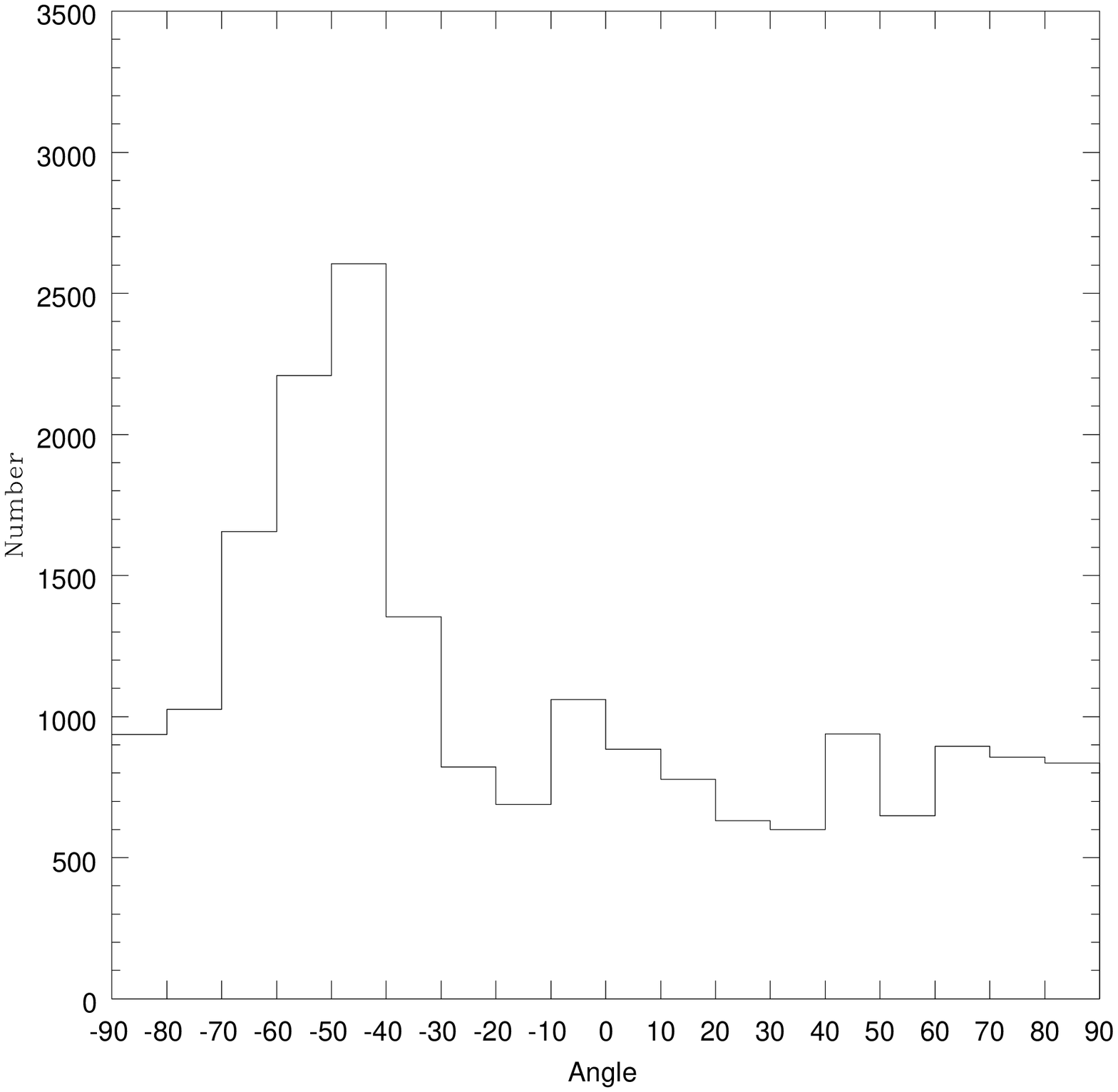,width=6cm}}
\Large
{\bf d)}
\normalsize
\subfigure{\psfig{figure=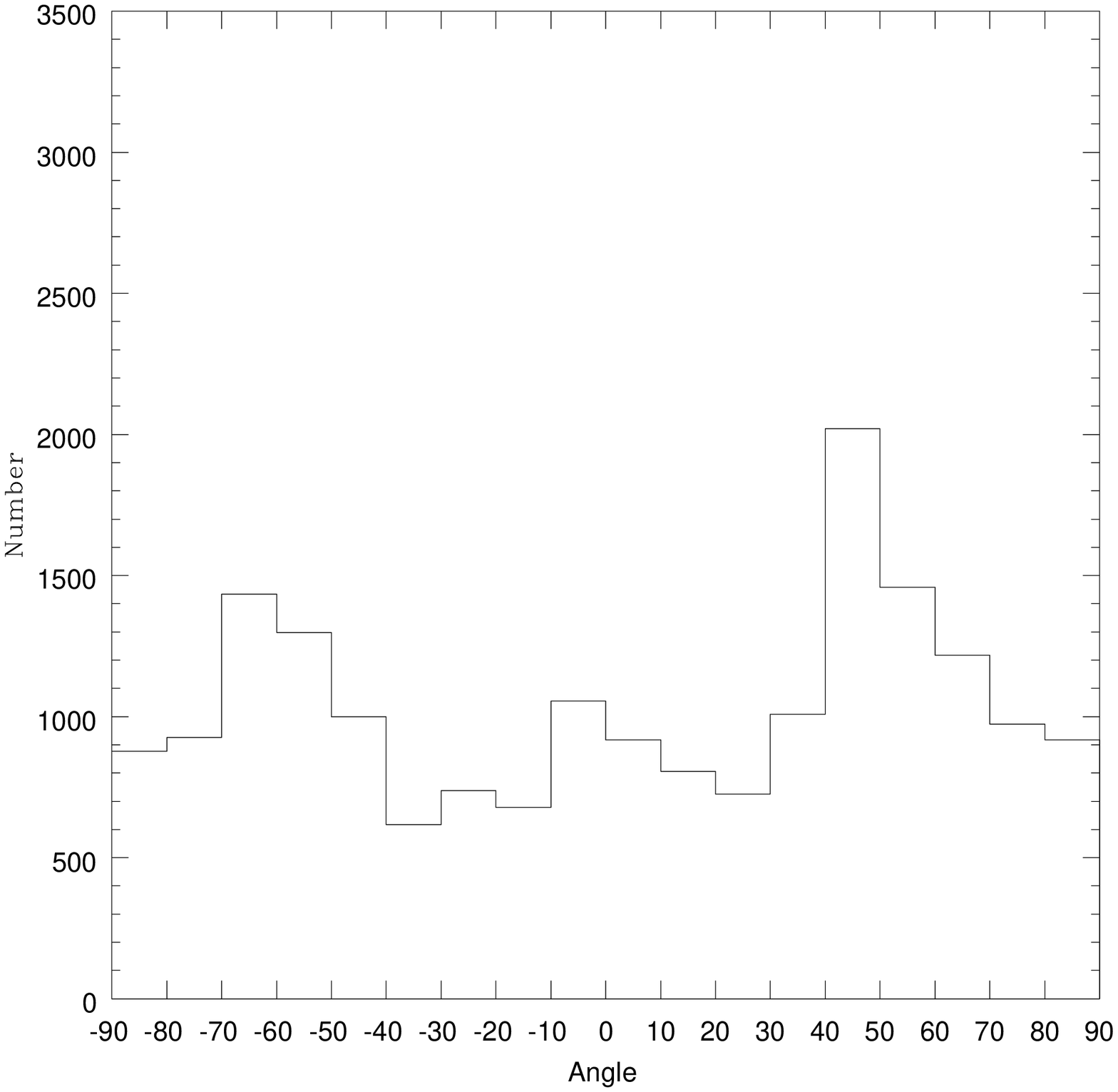,width=6cm}}

\Large
{\bf e)}
\normalsize
\subfigure{\psfig{figure=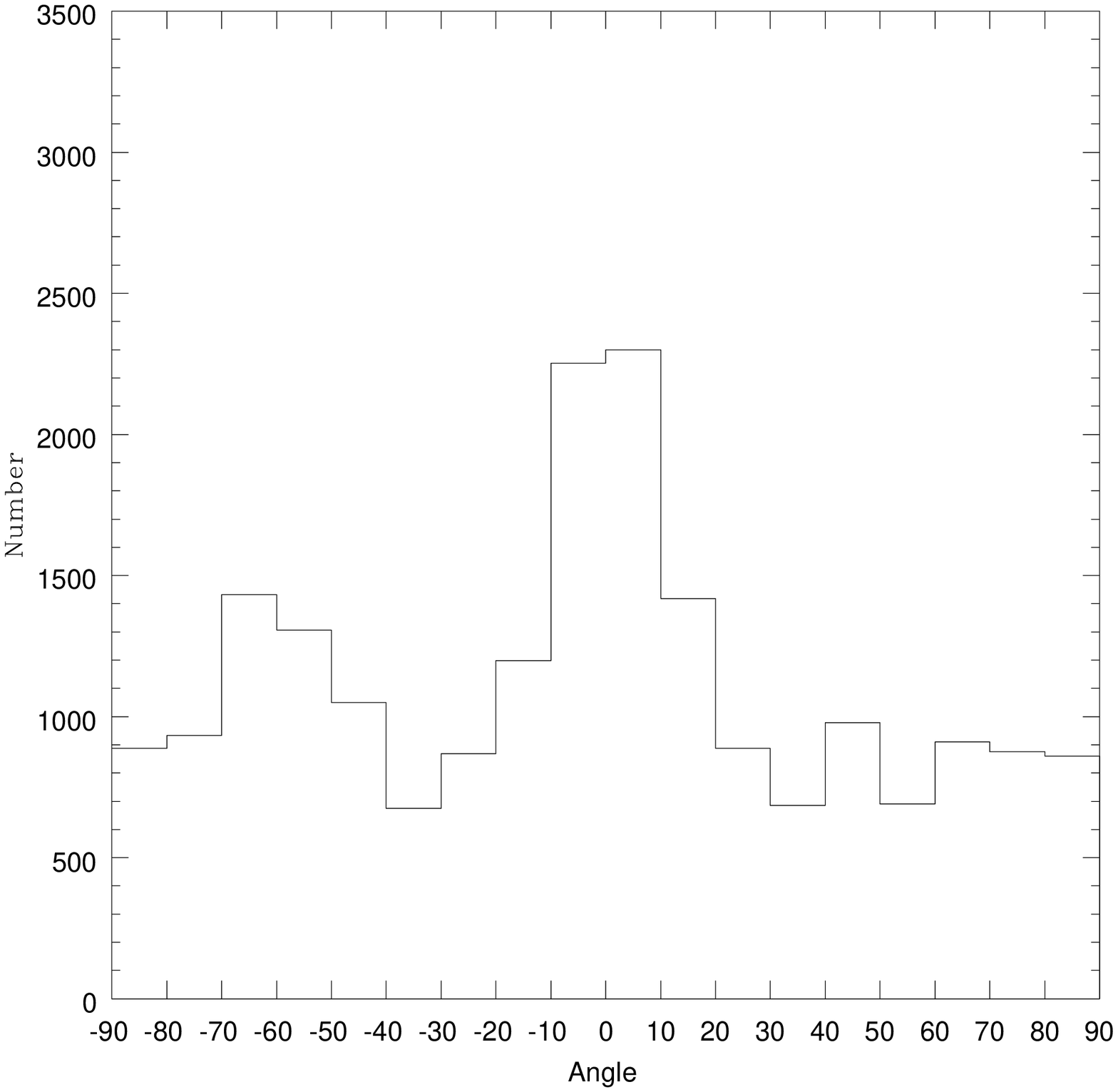,width=6cm}}
\Large
{\bf f)}
\normalsize
\subfigure{\psfig{figure=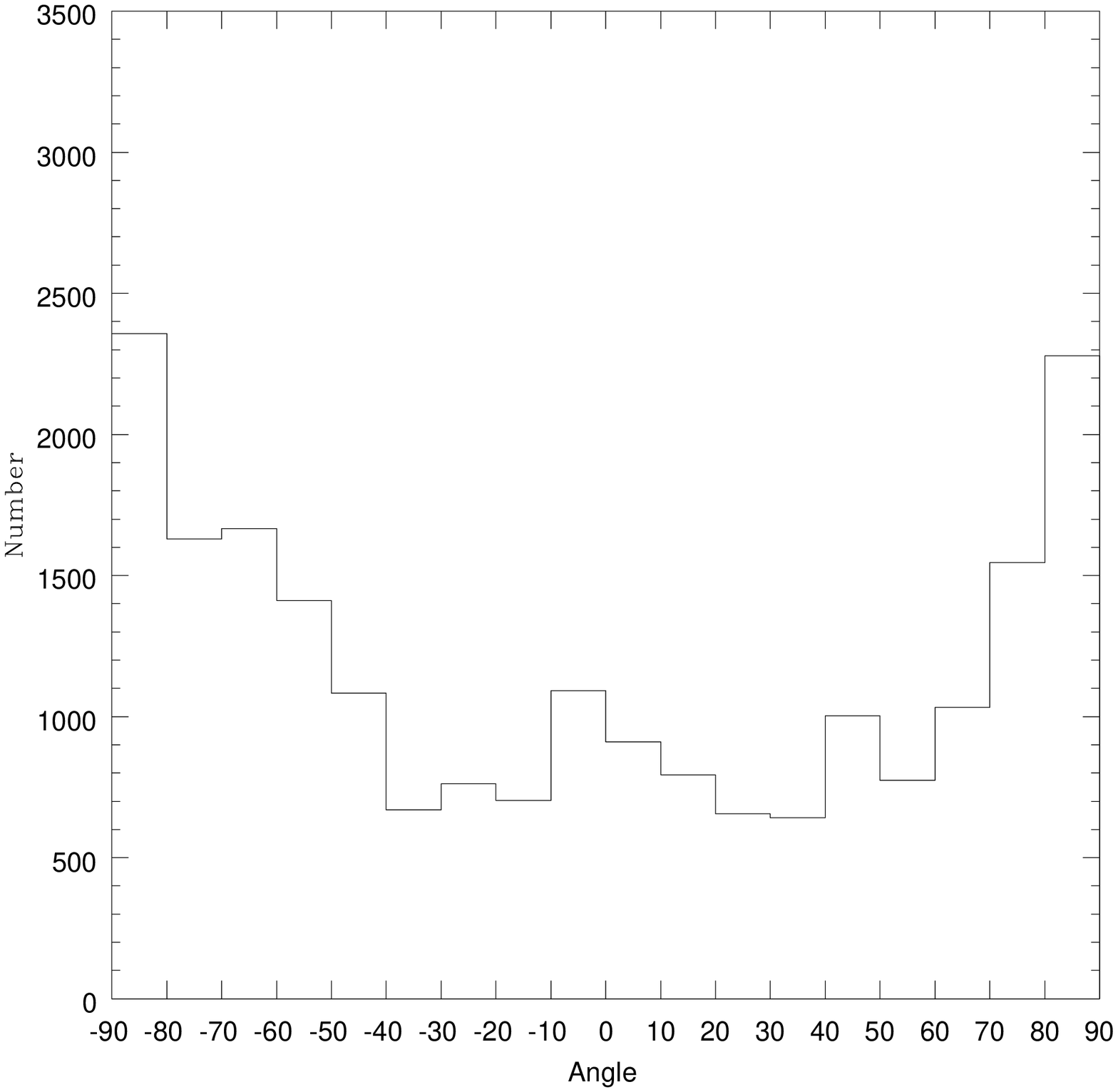,width=6cm}}
\caption
{Histograms of the angles that significant pixels make with the frame centre pixel. a) The smoothed data frame with no arcs added, b) Unsmoothed frame with an arc added, c) Smoothed with arc at an angle of -45 deg, d) Smoothed with an arc at an angle of 45 deg, e) Smoothed with arc at angle of 0 deg, f) smoothed with arc at an angle of 90 deg.}
\end{figure*}

We have applied the above detection method to our sample of 39 dE galaxies from the Virgo cluster. The result is that 26 (67\%) had $SN<2$ showing no sign of any tidal features. There were no galaxies with $SN>3$, and therefore no definite detections, although 13 (33\%) had $2<SN<3$ and so still remain possible detections.

\section{Moving to lower surface brightnesses}

One obvious critisism of the above technique is that the simulated arcs were added to data that was already reduced. This was done because the data is pipeline processed by the Cambridge Astronomical Survey Unit and then the reduced data is accessed from their archive. To go both deeper in surface brightness and to simulate the tidal streams in a more realistic way we have obtained additional observations of 7 galaxies from the Virgo cluster sample (Sabatini et al., 2003). 

In April 2004 we used the DOLORES instrument on the Galileo Telescope, La Palma (TNG) to obtain 50 minute exposures for each of our targets. The data was reduced in the normal way using the IRAF package and calibrated from photometric standards observed at regular intervals throughout the night (photometric accuracy to $\pm 0.1$ magnitudes). We chose to use the R band because with this instrumental set up we could go deeper in a given exposure time at R than at B. Typical early type galaxy colours are $1.0<(B-R)<1.5$ which we will use to compare the B band observations of the previous section with the R band observations discussed here.

To test the procedure described in the previous section we have added simulated arcs to the raw data (each image is made up of 5 separate 10 min exposures) and then processed the frame as before. Below we will show that an arc of surface brightness 27.5 R$\mu$ (28.5-29 B$\mu$) is easily detectable. This is 0.5 to 1.0 magnitudes deeper than the observations described in the previous section. In Fig. 3 we show the data frame (before smoothing) and the distribution of angles that low surface brightness pixels make with the origin. Looking carefully, the simulated stream can just be made out as a diagonal line across the CCD from bottom left to top right. In the histogram an arc at an angle of 45 degrees is clearly seen at $SN = 4.4$. It is quite clear that arcs at this surface brightness level are not removed by the data processing. Given the strong signal obtained for an arc at a surface brightness of 27.5 R$\mu$ we are confident that our procedure is capable of detecting arcs in this data set to an equivalent level of 29 B$\mu$.

\begin{figure*}
\subfigure{\psfig{figure=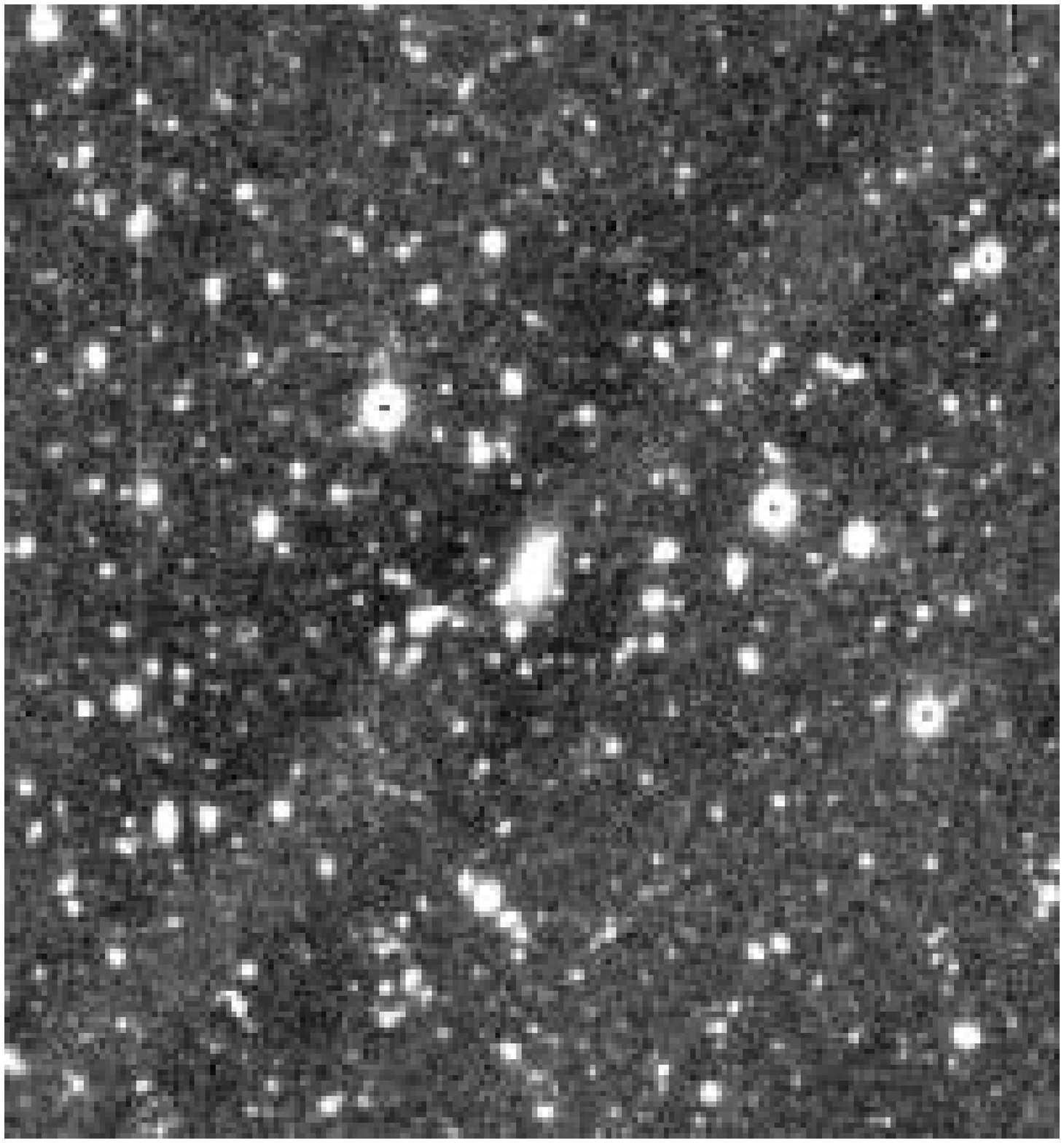,width=6cm}}
\subfigure{\psfig{figure=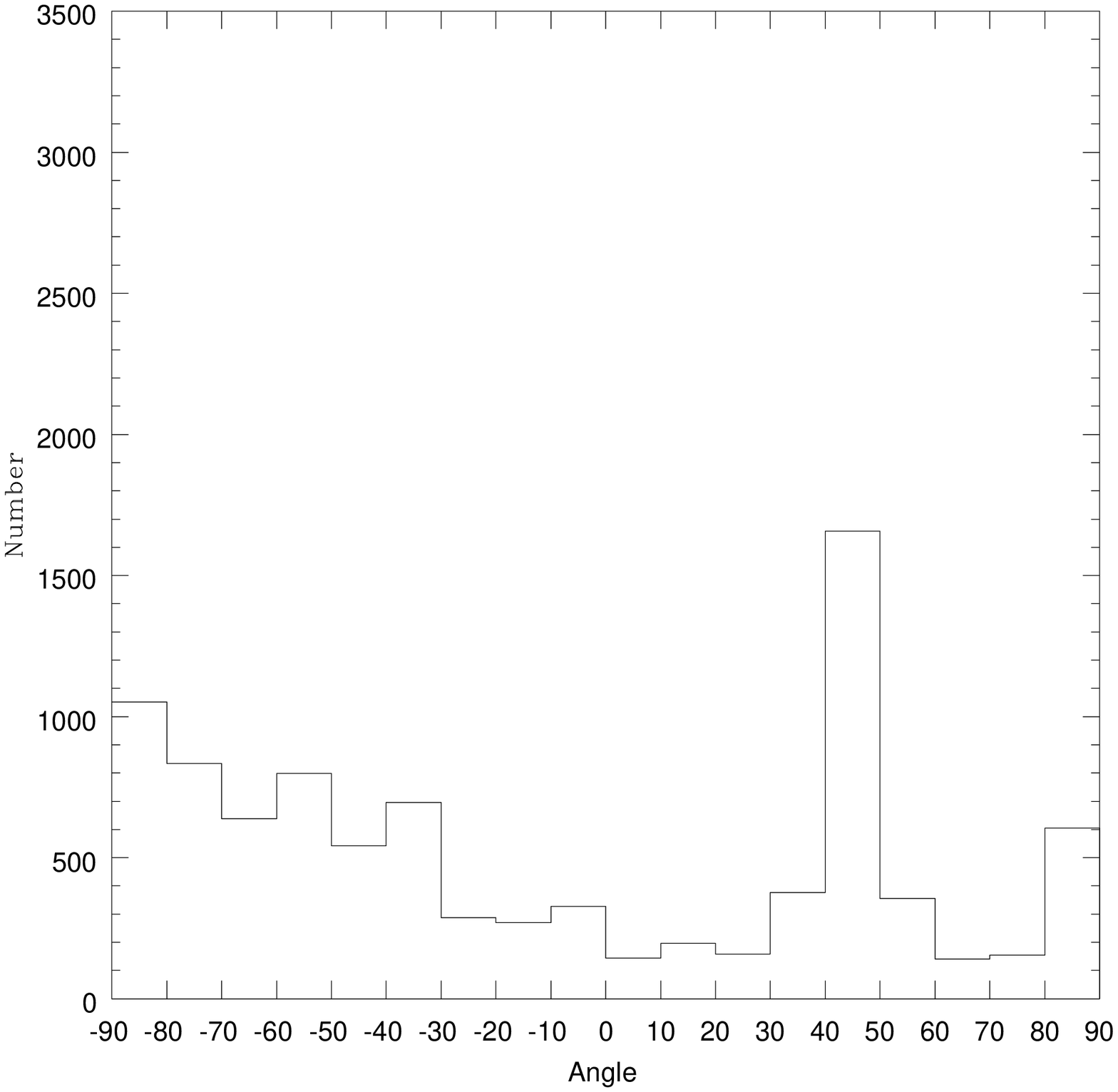,width=6cm}}
\caption
{In the longer exposure data a stream at a surface brightness level of 27.5 R$\mu$ ($28.5-29.0 B\mu$) is  just about visible. This simulated stream was added, with noise, to the raw data frames and then passed through the data reduction process. On the right we show the histogram of the angles that significant pixels make with the frame centre. There is a strong signal at 45 deg.}
\end{figure*}

We have applied the above procedure to the 7 galaxies in our sample. None of the galaxies have a detection at $SN>3$; one has $2<SN<3$. There is no evidence for any low surface brightness arcs.

\section{Discussion}
Although the original numerical simulations of galaxy harassment were carried out with a cluster designed to be like Coma, rather than Virgo, the conclusion was that this was a viable method for creating the excess of dwarf galaxies found in clusters in general (Moore et al., 1999).
The clear result from this paper is that we find little or no evidence for any extended tidal streams, with surface brightnesses less than 29 B$\mu$, associated with our target galaxies that would be the signature of morphological transformation by galaxy harassment. There are a number of possible explanations. The most obvious is that they do not exist (discussed in more detail below). The second is that they are too faint. This could arise because we have over estimated the stellar luminosity of the progenitor galaxies (the arcs are at much lower surface brightnesses) or because the tidal material was pulled out of the galaxies a long time ago (more than a few billion years) and has since dispersed. We cannot rule out either of these two, but we believe that there are other reasons for being skeptical about the  harassment model - the morphological transformation of LSB disc galaxies into dE galaxies:
\begin{enumerate}
\item There needs to be a large population of low surface brightness disc galaxies that reside in the Universe at redshifts of order $z=0.4$, awaiting morphological transformation.
\item Moore et al. (1996) quote a number of articles that clearly demonstrate the Butcher-Oemler effect - that is that the fraction of galaxies with colours similar to late type galaxies is higher in distant clusters compared to nearby clusters. They then associate this with the morphological transformation of galaxies. It is clear that galaxies are undergoing enhanced star formation and can be morphologically disturbed in the cluster environment, but the Butcher-Oemler effect is not evidence for morphological transformation.
\item There are many dE galaxies that are just too small to have been re-shaped by harassment. If the cluster dwarf galaxy population originated from a tidally truncated population of larger field galaxies then their tidal radii should be $\approx R_{c} \frac{\sigma_{dwarf}}{\sigma_{Clust}}$ where $R_{c}$ is the cluster core radius ($\approx0.5$ Mpc for Virgo), $\sigma_{dwarf}$ ($\approx 10$ km s$^{-1}$) is the internal velocity dispersion of the dwarf and $\sigma_{Clust}$ ($\approx 700$ km s$^{-1}$) is the  velocity dispersion of the cluster. The smallest tidal radius is of order 7 kpc. This is much larger than most of the dE galaxies we have in our Virgo cluster sample (radii of order 10 arc sec, less than 1 kpc). 
\end{enumerate}

We have previously proposed that dwarf galaxies are rather robust objects in the cluster environment because they have large mass-to-light ratios (Sabatini et al., 2004). However, in clusters like Virgo they do undergo accelerated evolution (star formation) due to tidal interactions in the cluster (see for example Henriksen \& Byrd, 1996) which eventually exhausts their gas supply. Thus the cluster environment is one that reveals its dark matter halos to us.
So, are cluster dwarf galaxies the embers of a once much brighter fire or are they ignited for the first time in the cluster environment ?

\end{document}